\documentclass[10pt]{iopart}
\usepackage{epsfig,amsfonts,cite}
\usepackage{pstricks}

\def\be{\begin{equation}}
\def\ee{\end{equation}}
\def\bea{\begin{eqnarray}}
\def\eea{\end{eqnarray}}
\def\bpar{\left(\!\!\begin{array}}
\def\epar{\end{array}\!\!\right)}
\def\bdar{\left|\!\!\begin{array}}
\def\edar{\end{array}\!\!\right|}
\def\bar{\begin{array}}
\def\ear{\end{array}}
\def\dsp{\displaystyle}

\def\e{{\rm e}}
\def\Fu{F_\uparrow}
\def\Gu{G_\uparrow}
\def\Fd{F_\downarrow}
\def\Gd{G_\downarrow}
\def\Pu{\psi_\uparrow}
\def\Pd{\psi_\downarrow}
\def\bsigma{{\mbox{\boldmath $\sigma$}}}
\def\bTheta{{\mbox{\boldmath $\Theta$}}}
\def\bsigmasmall{{\mbox{\boldmath\tiny $\sigma$}}}

\def\bmu{\mbox{\boldmath $\mu$}}

\def\mat#1{{{{%\sffamily \bfseries 
	\bf#1}}}}
\def\H{{\mat H}}
\def\V{{\mat V}}
\def\Un{{\textrm{\bf 1}}_{2\times 2}}

\def\<{\langle}
\def\>{\rangle}

\begin{document}
\jl{31}
\title{\sffamily \bfseries 
	Mesoscopic rings with Spin-Orbit interactions}

\author{Bertrand Berche$^{(a,b)}$, Christophe Chatelain$^{(a)}$ 
and Ernesto Medina$^{(b,c,a)}$}
\address{$^{(a)}$Statistical Physics Group, Institut Jean Lamour, 
UMR  CNRS No 7198,\\ 
Universit\'e Henri
Poincar\'e, Nancy 1,\\
B.P. 70239, F-54506 Vand\oe uvre les Nancy, France\\
$^{(b)}$Laboratorio de F\'isica Estad\'istica de Sistemas Desordenados, Centro de F\'\i sica, Instituto Venezolano de Investigaciones Cient\'\i ficas, \\
Apartado 21827, Caracas 1020 A, Venezuela \\
$^{(c)}$Escuela de F\'\i sica, Facultad de Ciencias, 
	Universidad Central de Venezuela,\\ 
	Caracas, Venezuela
}
\date{\today}
\begin{abstract}
\baselineskip=9pt
A didactic description of charge and spin equilibrium currents on mesoscopic rings in the
presence of Spin-Orbit interaction is presented.  Emphasis is made on 
the non trivial construction of the correct Hamiltonian in polar coordinates, the calculation of eigenvalues and 
eigenfunctions and the symmetries of the ground state properties. Spin currents are derived following an intuitive definition and then a more thorough derivation is built upon the canonical Lagrangian formulation that emphasizes the SU(2) gauge structure of the transport problem of spin $1/2$ fermions in spin-orbit active media. The quantization conditions that follow from the constraint of single-valued
Pauli spinors are also discussed. The targeted students are those of a graduate 
Condensed Matter Physics course.
\\
\\
{\bf R\'esum\'e.} 
On donne une description didactique des courants de charge et de spin dans un
anneau m\'esoscopique en pr\'esence de couplage spin-orbite. L'accent est mis
sur la construction de l'hamiltonien, le calcul des valeurs propres et des
vecteurs propres et les propri\'et\'es li\'ees \`a la sym\'etrie
par renversement temporel. Le courant de spin est calcul\'e \`a partir d'une 
d\'efinition intuitive, puis une d\'emonstration plus approfondie est 
construite \`a partir de la formulation lagrangienne qui met en \'evidence
la structure de jauge SU(2).  
 Les conditions de quantification qui d\'ecoulent 
des contraintes de raccordement des spineurs de Pauli sont \'egalement
discut\'ees. Le public vis\'e est celui d'un cours avanc\'e de physique de
la mati\`ere condens\'ee.
\end{abstract}
\ \hskip2.37cm\today

%\maketitle
 
%%%%%%%%% INTRODUCTION %%%%%%%%%

\section{Introduction}
\label{sec:intro}
The study of spin transport properties in materials, especially in 
semi-conductors where two-dimensional electron gases (2DEG) can be fabricated,
has become an important field of research both theoretically and experimentally
in the last decade. One of the key concept there is the Spin-Orbit (SO) 
interaction which allows for spin manipulations. Mesoscopic rings are  very simple laboratories for the investigation of
various quantum effects~\cite{Imry} in the presence of spin-orbit 
interactions. They  are the subject of intensive interest, both for fundamental reasons and potential 
applications in spintronics devices. Let us only mention a few fundamental studies in connection with phase effects,
such as Aharonov-Casher effect~\cite{Konig06,Kovalev07} or Sagnac
phase shifts~\cite{Zivkovic08}, spin interferences and spin 
filtering~\cite{Frustaglia04,Harmer06,Hatano07,Berche09,Lopez09} 
or in the fashionable subject of  graphene band structure~\cite{Recher07}.
On the more applied side, we can cite spin manipulations~\cite{Zhang06,Foldi06,Citro06,Nitta09}, studies
of spin related transport properties~\cite{Molnar04,Zhang07,Foldi09} 
and persistent currents~\cite{Splettstoesser03,Sun07,Huang09}, spin filtering in
arrays of SO quantum rings~\cite{Kalman08} that operate on principles of spin
interferometry analogous to their optical counterparts~\cite{Shelykh09} 
and concepts for the elaboration of qubits gates~\cite{Foldi05}. 

Many of the concepts involved in spin transport are rooted in very basic quantum
mechanics but in contexts that may be unfamiliar for the beginner, such as
multiply connected geometries like closed loops or rings. This paper will serve
to sort out such uncommon applications opening the scope of understanding
forefront topics such as spintronics. 

In this paper, we present an analysis of spin transport in a mesoscopic ring.
The exposition is suitable for a course in condensed matter physics at the 
graduate level. We carefully present the details of the non trivial construction of the
correct Hamiltonian in polar coordinates, the calculation of eigenvalues and 
eigenfunctions and the symmetries of the ground state properties with respect
to time reversal symmetry. Charge and spin  currents are derived following an intuitive 
and heuristic definition and then a more thorough derivation is built upon the canonical Lagrangian formulation 
that emphasizes the $SU(2)$ gauge structure of the transport problem. 
The path integral approach and the quantization conditions that follow 
from the constraint of single-valued Pauli spinors are also discussed for the benefit
of a full picture that will be valuable to the student. 

The Spin-Orbit interaction finds its origin in the Pauli
equation, which follows from the non-relativistic limit of the Dirac equation,
\be\bar{rcl}
\H&\!\!=\!\!&
\dsp\Bigl[\frac{({\vec p}-e{\vec {\rm A}})^2}{2m}
-e\phi\Bigr]\Un 
-\Bigl[\frac{{\vec p\ \!}^4}{8m^3c^2}
-\frac{e\hbar^2}{8m^2c^2}\vec\nabla\cdot\vec {\rm E}\Bigr]
\Un\\ \\
&&\dsp\quad-\frac{e\hbar}{2m}{\vec\bsigma}\cdot \vec{\rm B}
+\frac{e\hbar{\vec \bsigma}\cdot ({\vec p}-e\vec{\rm A})
\times\vec{\rm E}}{4m^2c^2}
.
\ear\label{PauliFirst}
\ee
Here we consider electrons, $e=-|e|$.
Ordinary vectors are denoted by arrows and bold face characters are used
for $2\times 2$ matrices. 
The first term in
the first line corresponds to the usual Schr\"odinger equation including
the kinetic energy with a mininal coupling  to the electromagnetic 
gauge field $\vec{\rm A}$
and a scalar potential contribution $-e\phi$.  
The second term in the first line describes the 
first relativistic correction to the kinetic energy and the Darwin term,
where $\vec{\rm E}$ is the electric field and $c$ the speed of light. These 
first two terms are proportional to the $2\times 2$ identity matrix
in spin space $\Un$. The second
line comprises explicitly spin-dependent terms, first the Zeeman
interaction\cite{Sakurai} where $\vec{\rm B}$ is the magnetic field and $\vec\bsigma$ 
is the vector of Pauli matrices\cite{Sakurai}
and the second term is the Spin-Orbit interaction,
written with the minimal coupling to the gauge vector. 
We have assumed a static potential so that the rotor of the electric field 
is absent and the Spin-Orbit interaction is limited to the term mentioned 
here~\cite{BjorkenDrell}.  

Forgetting about the magnetic field, the Spin-Orbit interaction has
a simple interpretation in terms of the interaction of the spin
magnetic moment of the particles, (here supposed to be electrons of 
Land\'e factor\cite{Merzbacher} $g\simeq 2$ and spin ${\vec {\mat s}}
=\frac 12\hbar{\vec \bsigma}$),
$\vec{\bmu}=g\frac{e}{2m}{\vec {\mat s}}=
-\frac{|e|\hbar}{2m}{\vec\bsigma}$, 
with the magnetic field produced by all
external moving charges in the electron rest frame.
Assuming a uniform electric field $\vec{\rm E}$ acting on the moving
electrons, the SO contribution to the Hamiltonian results from the interaction
$-\vec{\bmu}\cdot\vec{\rm B}_{\hbox{\small rest fr.}}$ of $\vec{\bmu}$ 
with the effective magnetic field
experienced by the particles in their rest frame. In the case of a Lorentz
change of reference frame, one has
\be \vec{\rm B}_{\hbox{\small rest fr.}}=\frac 1{c^2}(-\vec{v})\times\vec{\rm E}
=-(mc^2)^{-1}
({\vec p}\times\vec{\rm E}),\ee
where $\vec v$ and $\vec p$ refer to the dynamical variables of the electron.
This expression is corrected by a similar contribution with a factor
$-\frac 12$ due to the Thomas precession in the case of a closed orbit. Such a term
appears when the proper Lorentz transformation for the fields is considered\cite{Merzbacher}.  In 
such a way, the resulting SO interaction is usually written (e.g. in atoms)
\begin{eqnarray}
 H_{\rm SO}&=& -\frac{|e|\hbar}{2m^2c^2}
\vec\bsigma\cdot({\vec p}\times\vec{\rm E}),\nonumber\\
&=&\frac{|e|}{m^2c^2}\frac{1}{r}\frac{\partial \phi(r)}{\partial r}
\vec{\bf s}\cdot{\vec L},
\end{eqnarray}
were we have substituted a spherically symmetric potential, $\vec{\rm E}=-{\vec \nabla}\phi
= \frac{1}{r}\frac{\partial \phi(r)}{\partial r}{\vec r}$ and the interaction is proportional to 
${\vec \mat s}\cdot{\vec L}$. It is clear now why, in the context of atoms, the interaction is christened
the Spin-Orbit interaction i.e. ${\vec L}$ pertains to the angular momentum of the orbit while $\vec s$ pertains to intrinsic angular momentum, the spin.

\begin{figure} [th]
\vspace{0cm}
        \epsfxsize=7.0cm
        \begin{center}
        \mbox{\epsfbox{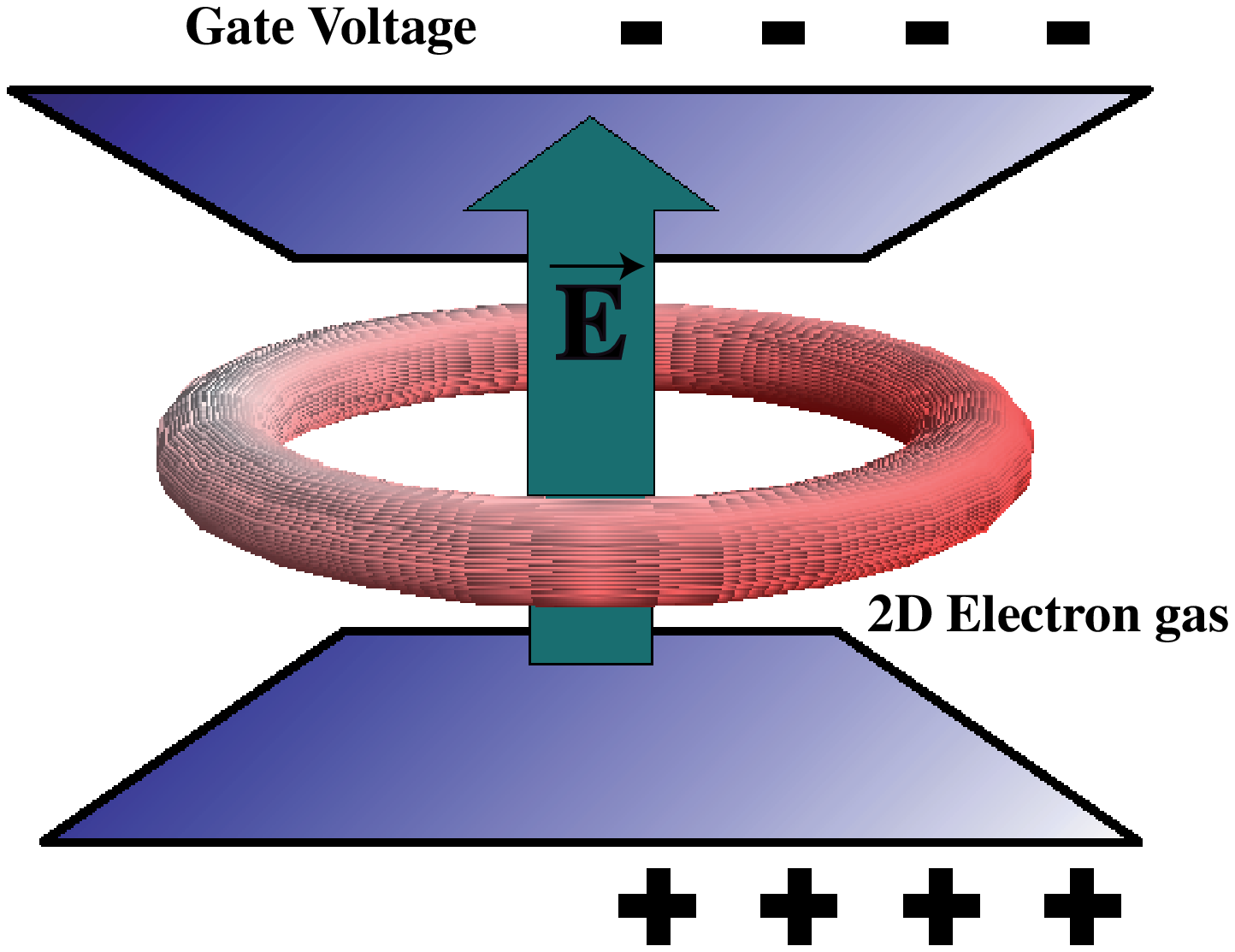}\qquad}
        \end{center}\vskip 0cm
        \caption{The typical Rashba situation with an electric field, created by a gate voltage,
	perpendicular to a ring fabricated on a two dimensional electron gas.}
        \label{FigRashba}  
\end{figure}

The prefactor in this expression depends on the details of the problem
(Land\'e factor, effective mass, \dots)  its sign even depends on the 
nature of the particles involved. In semiconductor physics, one usually introduces the notations $\alpha$ or
$\beta$ for  this coefficient, which is determined perturbatively in the ${\vec k}\cdot{\vec p}$ theory by a matrix element
in the Bloch wave functions basis. 
%The Rashba SO interaction then takes
%the following form 
In the case of a two-dimensional electron gas with
a gate voltage applied perpendicular to the $2d$ sample (with a non-symmetric confining potential generating
a space inversion assymmetry (SIA)),
this term is known as the Rashba SO interaction~\cite{Rashba,BychkovRashba},
\be
\V_{Rashba}={\hbox{const.}}\  \vec\bsigma\cdot(\vec{\rm E}\times{\vec p})=
\alpha (\bsigma_yp_x-\bsigma_xp_y).\label{eqRashba}
\ee
Note that the Rashba SO amplitude can be tuned experimentally using a gate
voltage, since the prefactor $\alpha$ is proportional to the electric
field (see figure~\ref{FigRashba}). 

The calculation of matrix elements leading to $\alpha$ 
is generally a hard task and one usually uses phenomenological expressions compatible with the
crystal symmetries or computes them directly from experiment\cite{Ganichev}. The Rashba interaction applies in the case of SIA. When there is bulk inversion asymmetry (BIA) i.e. the crystal unit cell lacks inversion symmetry, we have the Dresselhaus `flavor' of the SO interaction in $3d$ systems~\cite{EngelRashbaHalperin},
\begin{equation}
\V_{D,3d}={\rm const.}\ \!k_x(k_y^2-k_z^2)\bsigma_x+{\rm c.p.}
\end{equation}
where ${\rm c.p.}$ stands for cyclic permutations. In the case of electrons  
confined in two dimensions, the expectation value along the confinement, quantized
dimension should be considered. If that direction is $z$, then $\langle k_z\rangle\simeq 0$,
$\langle k_z^2\rangle\simeq(\pi/\ell)^2$, $\ell$ being the typical confinement 
length or spatial width of the confinement potential. The Dresselhaus SO interaction thus takes the simple form, here written
in a notation closer to equation~(\ref{eqRashba}),
\begin{equation}
\V_{Dresselhaus}=\beta (\bsigma_xp_x-\bsigma_yp_y),\label{eqDresselhaus}
\end{equation}
where we neglect cubic terms in $k$ and the average values computed are lumped in the definition of $\beta$. For more details on SO interactions in semi-conductors, see 
Refs.~\cite{EngelRashbaHalperin,winkler,Rashba06}.

\section{Derivation of the ring Hamiltonian}
\subsection{The argument of Meijer, Morpurgo and Klapwijk for the Rashba
SO interaction}
To make the discussion simpler, we neglect any magnetic field effect, which is well discussed in
the literature. Here we discuss the non-trivial point of designing a Hermitian Hamiltonian
when it is not in cartesian coordinates. We explain in detail such
construction so the student can understand the pitfalls generally avoided in quantum mechanics courses
because of a preferential system of coordinates. Many researchers overlooked this hermiticity
problem until the recent paper by Meijer, Morpurgo and Klapwijk~\cite{MMK02}.

The classical argument in reference \cite{MMK02} 
is essentially the following. We consider a 2DEG, neglecting interactions between electrons  
and add the Rashba SO interaction in 
cylindrical coordinates $(\rho,\varphi,z)$,
\be\fl\bar{rcl}
\H_{2d}&\!\!=\!\!&\dsp -\frac{\hbar^2}{2m}
\left(\partial_\rho^2+\rho^{-1}\partial_\rho+\rho^{-2}\partial_\varphi^2
\right)\Un\\ \\
&&+\alpha \rho^{-1}(\bsigma_x\cos\varphi+\bsigma_y\sin\varphi)
i\hbar\partial_\varphi
+\alpha (\bsigma_x\sin\varphi-\bsigma_y\cos\varphi)
i\hbar\partial_\rho.\\
\ear
\label{eqH2D}
\ee
%This Hamiltonian is a $2\times 2$ matrix. 
The Rashba SO interaction in the second line comes from 
equation~(\ref{eqRashba}) with the definitions $p_x=-i\hbar\partial_x$
and $p_y=-i\hbar\partial_y$
after the substitutions
$\partial_x=-(\rho)^{-1}\sin\varphi\ \!\partial_\varphi
+\cos\varphi\ \!\partial_\rho$
and 
$\partial_y=(\rho)^{-1}\cos\varphi\ \!\partial_\varphi
+\sin\varphi\ \!\partial_\rho$.
Fixing the radial distance $\rho$ to the ring radius $a$ and neglecting the
radial derivatives leads to 
a ``$1d$'' Hamiltonian (called ${\H'}_{1d}$ for
the moment) for the ring,
\be
{\H'}_{1d}=\frac{\hbar^2}{2ma^2}(i\partial_\varphi)^2+\alpha \hbar a^{-1}
(\bsigma_x\cos\varphi+\bsigma_y\sin\varphi)i\partial_\varphi.
\label{eqHPrime1d}\ee
In \cite{MMK02}, the authors note that the last term in (\ref{eqHPrime1d}) is 
non-Hermitian, since
\bea\fl
%\bar{rcl}
\<F|(\bsigma_x\cos\varphi+\bsigma_y\sin\varphi)i\partial_\varphi|G\>^*
= 
\left(
	\int_0^{2\pi}d\varphi\ \! (\Fu^*\ \Fd^*)
	\bpar{cc}
	0 & i\e^{-i\varphi}\partial_\varphi\\
	i\e^{i\varphi}\partial_\varphi & 0 
 	\epar
	\bpar{c} \Gu\\ \Gd\epar
\right)^*
\nonumber\\
\fl\qquad = 
\left[
	-i\Fu\e^{i\varphi}\Gd^*-i\Fd\e^{-i\varphi}\Gu^*
\right]_0^{2\pi}
+\int_0^{2\pi}d\varphi\ \! (\Gu^*\ \Gd^*)
	\bpar{cc}
	0 & i\e^{-i\varphi}\partial_\varphi\\
	i\e^{i\varphi}\partial_\varphi & 0 
 	\epar
	\bpar{c} \Fu\\ \Fd\epar
\nonumber\\
\fl\qquad\ \  
+\int_0^{2\pi}d\varphi\ \! (\Gu^*\ \Gd^*)
	\bpar{cc}
	0 & \e^{-i\varphi}\\
	-\e^{i\varphi} & 0 
 	\epar
	\bpar{c} \Fu\\ \Fd\epar.
\nonumber\\
\fl\qquad =
\<G|(\bsigma_x\cos\varphi+\bsigma_y\sin\varphi)i\partial_\varphi|F\>
-i\<G|(\bsigma_x\sin\varphi-\bsigma_y\cos\varphi)|F\>,
%\ear
\label{eqHermiticity}
\eea 
where $|F\>$ and $|G\>$ are Pauli spinors, 
and the integrated terms vanish if we impose
single-valued spinors, i.e. $\langle \varphi+2\pi|F\>=\<\varphi|F\>$. 
From the property given in equation (\ref{eqHermiticity}), we may infer that a
Hermitian operator could be formed by the combination
$\V=\alpha\hbar a^{-1}\mat W$ where
\be \mat W=(\bsigma_x\cos\varphi+\bsigma_y\sin\varphi)i\partial_\varphi
-iA(\bsigma_x\sin\varphi-\bsigma_y\cos\varphi),\label{eqW}
\ee
since 
\be
\bar{rcl}
 \<F|\mat W|G\>^*&\!\!=\!\!&
\<G|(\bsigma_x\cos\varphi+\bsigma_y\sin\varphi)i\partial_\varphi|F\>\\
&&-i(1-A^*)\<G|(\bsigma_x\sin\varphi-\bsigma_y\cos\varphi)|F\>,
\ear
\label{eqFWG}\ee
and
\be
\bar{rcl}
  \<G|\mat W|F\>&\!\!=\!\!&
\<G|(\bsigma_x\cos\varphi+\bsigma_y\sin\varphi)i\partial_\varphi|F\>\\
&&-iA\<G|(\bsigma_x\sin\varphi-\bsigma_y\cos\varphi)|F\>.\ear 
\label{eqGWF}\ee
The r.h.s. of equations (\ref{eqFWG}) and  (\ref{eqGWF}) are equal 
provided that 
$A^*=A=\frac 12$. This value  in equation (\ref{eqW})
gives the correct expression that renders the Rashba SO 
interaction on the ring Hermitian. 

The method used in Ref.~\cite{MMK02} to build hermiticity into the SO Hamiltonian,
consists in the introduction of a confining potential $V(\rho)$
in equation (\ref{eqH2D}) in order to localize the particle on a circle of radius $a$.
If this potential is steep enough, the particles will lie in the ground state
$R_0(\rho)$ of 
\be
H_{\rho}={\rm KE} + V(\rho) = -\frac{\hbar^2}{2m}
\left(\partial_\rho^2+\rho^{-1}\partial_\rho
\right)+V(\rho),
\ee
where ${\rm KE}$ is the contribution of the radial motion to the kinetic energy
and the correct $1d$ Hamiltonian 
($\H_{Rashba}^{\hbox{\footnotesize ring}}$) 
is defined by the projection,
on this ground state, of the angular dependence of $\H_{2d}$,
\be
\H_{Rashba}^{\hbox{\footnotesize ring}}=\<R_0|{\H}_{2d}
%+\frac{\hbar^2}{2m}\left(\partial_\rho^2+\rho^{-1}\partial_\rho\right)
-{\rm KE}\ \!\Un|R_0\>,
\ee
where $|R_0\>$ is extended to a two-component object. Note the
very peculiar situation in polar coordinates, ${\rm KE}$
does not simply coincide with $p_\rho^2/2m$ (see e.g. Ref.~\cite{Paz} for
a discussion on this unrecognized point).  
The calculation is now made specific by considering a circular Harmonic 
potential $V(\rho)=\frac 12m\omega^2(\rho-a)^2$ and by the explicit 
calculation of the matrix 
elements $\<R_0|\rho^{-1}|R_0\>=a^{-1}$ and $\<R_0|\partial_\rho|R_0\>=
-(2a)^{-1}$. The resulting Hamiltonian $\H_{Rashba}^{\hbox{\footnotesize ring}}$ 
then follows from
the substitutions $\rho\to a$, $\partial_\rho\to-(2a)^{-1}$ 
in ${\H}_{2d}+\frac{\hbar^2}{2m}
\left(\partial_\rho^2+\rho^{-1}\partial_\rho
\right)$. Thus
\be\fl
\bar{rcl}
\H_{Rashba}^{\hbox{\footnotesize ring}}
&\!\!=\!\!&\dsp \frac{\hbar^2}{2ma^2}(i\partial_\varphi)^2
\Un\\ \\
&&+\alpha \hbar a^{-1}(\bsigma_x\cos\varphi+\bsigma_y\sin\varphi)
i\partial_\varphi
-i\alpha \hbar (2a)^{-1}(\bsigma_x\sin\varphi-\bsigma_y\cos\varphi).\ear
\label{eqH1d}
\ee
It can be checked that the Hamiltonian is now Hermitian, since it coincides 
with the choice $A^*=A=\frac 12$, according to the procedure outlined in 
Eqs.~(\ref{eqW}-\ref{eqGWF}).

%\begin{figure} [th]
%\vspace{0cm}
%        \epsfxsize=5.0cm
%        \begin{center}
%        \mbox{\epsfbox{Fig1.eps}\qquad}
%        \end{center}\vskip 0cm
%        \caption{Confining potentiel $V(\rho)$ on a ring of radius $a$.}
%        \label{Fig1}  
%\end{figure}

\subsection{Symmetrization of the original Hamiltonian}
Although the previous derivation is of course correct, it is instructive to 
propose another, probably more straightforward derivation. 
The SO interaction in cylindrical coordinates with a perpendicular
electric field, $\vec \alpha=\alpha {\vec e}_z$, takes the form
\be
\vec\bsigma\cdot({\vec\alpha}\times{\vec p})=-\alpha(\bsigma_\rho p_\varphi
-\bsigma_\varphi p_\rho),\label{beforesym}\ee
with $\bsigma_\rho=\bsigma_x\cos\varphi+\bsigma_y\sin\varphi$ and
$\bsigma_\varphi=-\bsigma_x\sin\varphi+\bsigma_y\cos\varphi$ 
treated for the moment as classical variables (i.e. we forget about 
possible commutation problems).
In order to make the corresponding expression Hermitian in Quantum Mechanics,
one has to use the correspondence principle, i.e. add inequivalent orders of non-commuting
operators, and symmetrize the ``classical''
expression, using also
$p_\rho=-i\hbar(\partial_\rho-(2\rho)^{-1})$ and 
$p_\varphi=-i\hbar a^{-1}\partial_\varphi$. Let us also 
note that when the electrons are
confined on a ring, then $\<p_\rho\>=\<R_0|p_\rho|R_0\>=0$
({\em and not} $\<R_0|\partial_\rho|R_0\>=0$ 
as it was done to get equation~(\ref{eqHPrime1d})) 
and $\rho = a$.
This leads to
\be
\V_{Rashba}^{\hbox{\footnotesize ring}} = -\frac 12\alpha\{\bsigma_\rho, p_\varphi\}=
i\hbar\alpha a^{-1}(\bsigma_\rho\partial_\varphi
+\frac 12\partial_\varphi\bsigma_\rho),
\ee
where $\{\bsigma_\rho, p_\varphi\}$ denotes the anticommutator
$\bsigma_\rho p_\varphi+ p_\varphi\bsigma_\rho$.
Using $\partial_\varphi\bsigma_\rho=\bsigma_\varphi$, equation~(\ref{eqH1d}) 
eventually follows,
\be
\H_{Rashba}^{\hbox{\footnotesize ring}}
=\frac{\hbar^2}{2ma^2}(i\partial_\varphi)^2\Un
+i\hbar\alpha  a^{-1}(\bsigma_\rho\partial_\varphi+\frac 12\bsigma_\varphi).
\label{eqRashbaRing}
\ee
Let us emphasize on the fact that the condition 
$\<p_\rho\>=0$ is equivalent to Meijer et al's derivation, since it implies
$-i\hbar\<R_0|\partial_\rho-(2\rho)^{-1}|R_0\>=0$ with a radial wave function
strongly localized on $\rho\simeq a$. Our proposed derivation, nevertheless, has the 
merit of being simpler and more general, since it does not rely
on any particular assumption on the confining potential (except the fact that 
it enforces the particles to move along a ring).

\subsection{Hamiltonian in the rotated spin basis}
In the previous sections, the Rashba Hamiltonian~(\ref{eqRashbaRing}), 
initially expressed in Cartesian coordinates, has been rewritten in polar coordinates.
To preserve the hermiticity of the Hamiltonian, it has been shown that an extra
term should be added if the expression is not properly symmetrized prior to 
quantization. In the following, we give a more physical picture
that gives rise to this extra term.

We first start by noting that the change of coordinates $(x,y)\rightarrow
(\rho,\varphi)$ induces a rotation by an angle $\varphi$ of the local frame:
$({\vec e}_x,{\vec e}_y)\rightarrow ({\vec e}_\rho,{\vec e}_\varphi)$.
All vector fields, including $\bsigma$ are now expressed
in this local frame. However, {\em no such
rotation} has been applied to the spinor basis which thus remains the same
at any point of the circle. The notations $\bsigma_\rho$ and $\bsigma_\varphi$
may thus be misleading. Let us introduce such a rotation of the spinor basis:
\be
|F'\>=e^{i{\varphi\over 2}\bsigmasmall_z}|F\>
=\bigl(\cos{\varphi\over 2}\Un+i\sin{\varphi\over 2}\bsigma_z\bigr)|F\>.
\ee
Since the rotation axis is $(Oz)$, the quantification axis is not changed
during the transformation. The two spin components simply get a phase
\be
|\uparrow'\>=e^{i{\varphi\over 2}}|\uparrow\>,\quad
|\downarrow'\>=e^{-i{\varphi\over 2}}|\downarrow\>.\quad
\ee
In the new local spinor basis, the Hamiltonian becomes
\be \H'=e^{i{\varphi\over 2}\bsigmasmall_z}\H 
e^{-i{\varphi\over 2}\bsigmasmall_z}.\ee
Using the identity
$ \bsigma_\rho\bsigma_z=-i\bsigma_\varphi$,
the correct Hermitian Rashba Hamiltonian can be cast as 
\be
\H_{Rashba}^{\hbox{\footnotesize ring}}
=\frac{\hbar^2}{2ma^2}(i\partial_\varphi)^2\Un
+i\hbar\alpha  a^{-1}\bsigma_\rho(\partial_\varphi
+\frac i2\bsigma_z).
\ee
Under the change of spinor basis, the new rotated Hamiltonian becomes
\be\fl
e^{i{\varphi\over 2}\bsigmasmall_z}\H_{Rashba}^{\hbox{\footnotesize ring}}
e^{-i{\varphi\over 2}\bsigmasmall_z}
=\frac{\hbar^2}{2ma^2}(i\partial_\varphi\Un+\frac 12\bsigma_z)^2
+i\hbar\alpha  a^{-1}\bsigma_x\partial_\varphi,
\label{eqRashbaRing2}
\ee
since $\bsigma'_\rho\equiv e^{i{\varphi\over 2}\bsigmasmall_z}
\bsigma_\rho e^{-i{\varphi\over 2}\bsigmasmall_z}=\bsigma_x$. Interestingly, after 
rotation of the spinor basis the potential energy recovers a form 
$\propto \bsigma'_\rho p_\varphi$ similar to the original Rashba Hamiltonian {\it before symmetrization} in~(\ref{beforesym}). From the physical point of view, one
indeed expects the potential energy to be the same in all local orthonormal
frames. This implies that, starting from the original Rashba Hamiltonian,
the change of coordinates should be accompanied by a change of the local
vector and spinor frames. The spin-orbit term then keeps the same form. When
expressed in the original spin basis, the additional term needed 
for symmetrization appears naturally.
In contradistinction to the potential energy, additional terms may
appear in the kinetic energy when one changes the spinor basis. They are interpreted as inertial
forces due to the fact that the local frame is not inertial.
This is the origin of the extra term $\frac 12\bsigma_z$ in the
Hamiltonian (\ref{eqRashbaRing2}) (note the appearance of the $z$ component of the total angular momentum, 
$\mat L_z+\mat s_z$). When the particle moves along the
circle, the local spinor basis turns by the same angle. Consider
for example a constant spinor. To accommodate the change of local
basis along the circle, it should have $\varphi$-dependent spin-components.
Since no kinetic energy is associated to the spin, a term involving 
derivatives $\partial_\varphi$ would produce a term which is 
precisely cancelled by the additional contribution $\frac 12\bsigma_z$.

\section{Eigenenergies and Eigenvectors on a ring}
\subsection{The Rashba SO interaction}

Once the correct Hermitian Hamiltonian has been written in polar 
coordinates we can confidently derive the eigenvalues and eigenvectors for the strictly one 
dimensional ring. This section will benefit the student by explicitly obtaining the
eigenfunctions and eigenvalues of the system, from which we can obtain all observables
or actually measurable quantities in experiments. We will also make some basic symmetry considerations
that will illuminate details of why the wavefunction and energies take their particular form.

First we rewrite the Hamiltonian in a clever way in order to arrive rapidly at 
the eigenvalues and introduce physical insight by pointing out the existence of a gauge field associated 
to topological phenomena connected to spin transport. An alternative form of the Hamiltonian~(\ref{eqRashbaRing}) 
was given in
Ref.~\cite{MolnarPeetersVasilopoulos}. We observe that
\be
\fl	\left(
	i\partial_\varphi\Un+\frac{ma\alpha}{\hbar}\bsigma_\rho
	\right)^2
	= (i\partial_\varphi)^2\Un
	+\frac{2ma\alpha}{\hbar}\bsigma_\rho (i\partial_\varphi)
	+\frac{ma\alpha}{\hbar}i\bsigma_\varphi
	+\left(\frac{ma\alpha}{\hbar}\right)^2\Un.
\ee
The factor of two in the second term on the right accounts for the appropriate action of an derivative operator. Such form allows one to 
recast the Hamiltonian as
\be
	\H_{Rashba}^{\hbox{\footnotesize ring}}
	=\frac{\hbar^2}{2ma^2}\left[\left(i\partial_\varphi
	\Un+\frac{ma\alpha}{\hbar}\bsigma_\rho
	\right)^2-\left(\frac{ma\alpha}{\hbar}\right)^2\Un\right].
\label{eqRashbaRingSquare}
\ee

In order to find the eigenstates of the 
Hamiltonian~(\ref{eqRashbaRingSquare}), one first solves the eigenvalue
equation
\be
	\left(i\partial_\varphi
	\Un+\frac{ma\alpha}{\hbar}\bsigma_\rho\right)\Psi=\varepsilon\Psi,
\ee
or
\be
\bpar{cc}
i\partial_\varphi & \frac{ma\alpha}{\hbar}\e^{-i\varphi}\\
\frac{ma\alpha}{\hbar}\e^{i\varphi}& i\partial_\varphi 
\epar
\bpar{c}
\Pu \\ \Pd
\epar
=\varepsilon
\bpar{c}
\Pu \\ \Pd
\epar.
\label{eigenvalueequation}
\ee
It follows that the general form for the eigenfuctions should be
\be
%\bpar{c}
%\Pu \\ \Pd
\Psi_{n,s}^{\lambda}=\e^{i \lambda n\varphi}
\bpar{c}
A_{\lambda, s}\e^{-i\varphi/2} \\ B_{\lambda,s}\e^{i\varphi/2}
\epar,
\label{eigenfunction}
\ee
\begin{figure} [th]
\vspace{0cm}
        \epsfxsize=12.0cm
        \begin{center}
        \mbox{\epsfbox{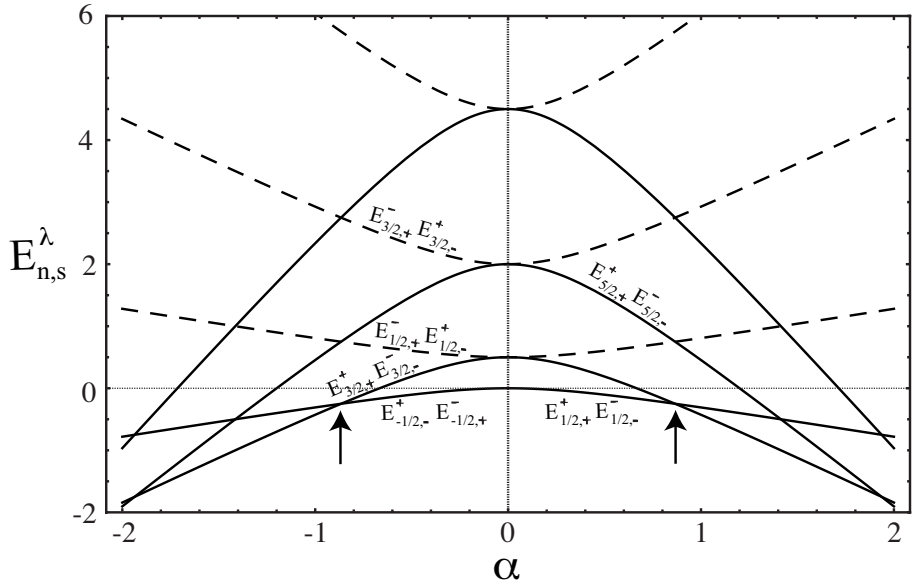}\qquad}
        \end{center}\vskip 0cm
        \caption{Energies of electrons on a ring with spin-orbit interaction 
as a function of the SO interaction strength $\alpha$. The energy is in units 
of $\hbar^2/ma^2$ while $\alpha$ is in units of $\hbar/ma$. Indicated are the 
eigenvalue labels for the first five levels (note the ground state is fourfold degenerate). The graph shows how the limit $\alpha=0$ is fourfold degenerate as expected since the time inversion symmetry turns into independent space and spin inversion symmetry. Arrows indicate change in lowest energy level.}
        \label{Fig1}  
\end{figure}
where $n$ is the main quantum number and we account for the two eigenvalues that will arise, $s=\pm$, for each wave propagation direction $\lambda=\pm$. So, assuming such a wavefunction we obtain the corresponding eigenvalues
\be
\varepsilon_{n,s}^{\lambda}=\frac{s}{2}\sqrt{1+4\left(\frac{ma\alpha}{\hbar}\right)^2}
-\lambda n, 
\ee
where $n$ is a half odd integer as will be seen below when we address the wavefunctions. The eigenvalues of (\ref{eqRashbaRingSquare}) then simply follow
\be
E_{n,s}^{\lambda}=\frac{\hbar^2}{2ma^2}\left[\left ( n-\frac{\lambda s}{2}\sqrt{1+4\left(\frac{ma\alpha}{\hbar}~\right)^2}
\right )^2-
\left(\frac{ma\alpha}{\hbar}\right)^2\right].\label{energies}
\ee
Note that left and right propagating waves with the same $s$-index are not 
degenerate but time reversal symmetry is satisfied i.e. 
$E_{n,s}^{\lambda}=E_{n,-s}^{-\lambda}$ (simultaneous change of $\lambda$ and $s$). This symmetry reflects the fact that the SO interaction is not space inversion symmetric, but only time reversal symmetric. When the spin-orbit interaction is absent ($\alpha=0$), space and spin inversion symmetry is recovered and energies of both clockwise and counterclockwise modes are degenerate independent of the spin label.  Figure \ref{Fig1} shows 
the energy levels as a function of the SO strength. The ordering of the 
levels are indicated according to 
the spin orientation and sense of the current. The free electron on a ring is recovered for all levels as one can see from the fourfold degeneracy at $\alpha=0$, rendering all possible combinations of the values of $\lambda,s$ for a fixed integer value of $N$ in the relation $E_{n,s}^{\lambda}=\hbar^2 (n-\lambda s/2)^2/2ma^2=\hbar^2 N^2/2ma^2$ (with $n$ half odd integers). 
Table 1 shows the energies for the first three levels in the limit of $\alpha=0$ and how degeneracies are broken when the SO interaction is turned on.

\begin{table}[ht]
\caption{Energies (in units of $\hbar^2/2ma^2$) for the first three levels and their degeneracies according to values of the quantum numbers $n$, $\lambda$ and $s$ in the limit of zero spin orbit, and the corresponding values of the free electron integer quantum number $N$. Also shown is how degeneracies occur when the SO interaction is turned on leaving only time reversal symmetry.}
\centering
    \begin{tabular}{  c  c  l  l  l  c }
    \br
    $E$ ($\hbar^2/2ma^2$) & $n$ & $\lambda$ & $s$ & $N$ & $\alpha\ne 0$ \\ 
	\mr
    0 & $-1/2$ & $+$ & $-$ & $0$ & deg \\ \cline{1-5}
    0 & $-1/2$ & $-$ & $+$ & $0$ &       \\ \cline{1-5}
    0 & $\phantom{-}1/2$  & $+$ & $+$ & $0$ &       \\ \cline{1-5}
    0 & $\phantom{-}1/2$  & $-$ & $-$ & $0$ &         \\ \mr
    1 & $\phantom{-}1/2$  & $+$ & $-$ & $1$ & deg \\ \cline{1-5}
    1 & $\phantom{-}1/2$  & $-$ & $+$ & $1$ &         \\ \mr
    1 & $\phantom{-}3/2$  & $+$ & $+$ & $1$ & deg \\ \cline{1-5}
    1 & $\phantom{-}3/2$  & $-$ & $-$ & $1$ &           \\  \mr
    4 & $\phantom{-}3/2$  & $+$ & $-$ & $2$ & deg \\  \cline{1-5}
    4 & $\phantom{-}3/2$  & $-$ & $+$ & $2$ &         \\  \mr
    4 & $\phantom{-}5/2$  & $+$ & $+$ & $2$ & deg \\  \cline{1-5}
    4 & $\phantom{-}5/2$  & $-$ & $-$ & $2$ &          \\  
	\br
    \end{tabular}
    \label{Table1}
\end{table}

Now we compute the eigenfunctions. In order to deal with the subtleties 
associated with the change in sign of spin and propagation direction we 
explicitly do the $\lambda=+$ case.  Using equations (\ref{eigenvalueequation}) and (\ref{eigenfunction}) we find from the secular equation
\begin{equation}
B_{+,s}=\frac{\hbar}{2ma\alpha}\left (\frac{s}{\cos{\theta}}-1\right )A_{+,s},
\end{equation}
where $\cos\theta=1/\sqrt{1+4(m a \alpha  /\hbar)^2}$. In order to conform 
to a canonical spinor we choose $A_{+,+}=\cos\theta/2$, and thus 
$B_{+,+}=\sin\theta/2$ for normalization, so we have
\begin{equation}
\frac{\hbar}{2 m a\alpha}\left (\frac{1-\cos\theta}{\cos\theta}\right ) 
\cos\frac{\theta}{2}=\sin\frac{\theta}{2},
\end{equation}
from which we have the condition
\begin{equation}
\tan\theta=\frac{2ma\alpha}{\hbar}.
\end{equation}
The choice for the second eigenfunction is $A_{+,-}=-\sin\theta/2$, and doing 
the same exercise we
arrive at the two eigenspinors
\begin{equation}
\Psi_{n,+}^+=e^{i n\varphi}\bpar{c}
\cos\frac{\theta}{2}\e^{-i\varphi/2} \\ \sin\frac{\theta}{2}\e^{i\varphi/2}
\epar,\\ \\
\Psi_{n,-}^+=e^{in\varphi}\bpar{c}
-\sin\frac{\theta}{2}\e^{-i\varphi/2} \\ \cos\frac{\theta}{2}\e^{i\varphi/2}.
\epar
\end{equation}
The corresponding eigenfunctions for $\lambda=-$ are
\begin{equation}
\Psi_{n,+}^-=e^{-in\varphi}\bpar{c}
\cos\frac{\theta}{2}\e^{-i\varphi/2} \\ \sin\frac{\theta}{2}\e^{i\varphi/2}
\epar,\\ \\
\Psi_{n,-}^-=e^{-in\varphi}\bpar{c}
-\sin\frac{\theta}{2}\e^{-i\varphi/2} \\ \cos\frac{\theta}{2}\e^{i\varphi/2}
\epar
\end{equation}
Using the time reversal operator for spin $1/2$ particles 
$\bTheta=-i \bsigma_y K$ where $\bsigma_y$ is the corresponding Pauli matrix 
and $K$ is the conjugation operator~\cite{PeskinSchroder}, 
one can readily show that 
$\Psi_{n,+}^+=\bTheta \Psi_{n,-}^-$ and $\Psi_{n,-}^+=\bTheta \Psi_{n,+}^-$. 
As one would expect from the time reversal invariance $\Psi_{n,+}^-$ is 
degenerate with $\Psi_{n,-}^+$ and $\Psi_{n,-}^-$ with $\Psi_{n,+}^+$. With these properties
in mind the student can guess that expectation values taken with corresponding time reversed
wavefunctions must be added together since there energies are degenerate. This
fact will have surprising consequences in the next section.

\section{Charge and spin currents in the ground state}
\subsection{Direct calculation}
Having found the eigenfunctions, one can now compute equilibrium properties 
such as persistent charge and spin currents on the ring. The student might be accustomed
to expecting a current only when an external force is applied to the system, such as a
potential difference. Nevertheless we will show that when symmetry breaking fields
exist that can do no work, such as a magnetic field, currents can exist as equilibrium properties
borne from the nature of the wavefunctions. In a sense the system can distinguish between
clockwise moving current and a counterclockwise moving current, one of them corresponding to a lower energy. These currents are long lived in the absence
of perturbations. This is very counterintuitive for the classical
line of thought.

There are charge currents in equilibrium only when time reversal symmetry breaking perturbations are present, such as a magnetic field\cite{PersistentCurrentOriginal,Imry}, for an experimental review see \cite{Mohanty} and references therein. The charge currents can be derived directly from 
the ground state energy as
\begin{equation}
I_{charge}=-\sum_{n,s}\frac{\partial E_{n,s}^{\lambda}}{\partial \Phi}
\end{equation}
where $\Phi$ is the magnetic flux and the $\lambda$ quantum number is chosen by the direction of the magnetic 
field. An alternative form of computing the currents is by using the 
definition of the current operator. These current densities are fields 
usually defined from a continuity equation rather than from matrix elements 
of some operators. In order to generalize the classical definition of the 
current density $\vec j(\vec r)=n(\vec r)e\vec v$, we take the expectation value of the 
velocity operator
\begin{equation}
\vec J_{charge} =\Psi^{\dagger}
	e \vec {\mat v}~\Psi.\\
\end{equation}
where $e$ is the electron charge and $\vec {\mat v}$ is the velocity operator.
Note that the current is defined in such a way that its
dimension is a charge times a velocity. 
In the presence of the spin-orbit interaction, the velocity operator is not simply $\vec {\mat p}/m$. There arises what is known as an additional anomalous velocity term. We start from the quantum mechanical definition of the 
velocity (which is now a $2\times 2$ matrix due to the presence of the 
spin-dependent terms in the Hamiltonian) $\vec {\mat v}
= \frac{i}{\hbar}[\H,\vec r ]$. Since we are interested in the azimuthal 
velocity component $\dot\varphi$,
it is more convenient to calculate
\begin{equation}
{\mat v}_{\varphi}=\frac{ia}{\hbar}[\H,\varphi]
=\frac{\hbar}{ima}\partial_{\varphi}\Un-\alpha\bsigma_{\rho}.
\label{velocityoperator}\end{equation}
rather than working in Cartesian coordinates. Note that the Hamiltonian
takes a simple form when expressed in terms of the velocity: 
${\mat H}=\frac 12 m({\mat v}_{\varphi}^2-\alpha^2\Un)$.
Here, we do not consider symmetry breaking magnetic fields, so there can be no persistent 
charge currents on the ring. If we compute the charge currents for the fourfold degenerate lowest energy levels from which
one forms a totally symmetric linear combination,
one gets  
\begin{eqnarray}
{J}_{\varphi}&=&-\frac e4 (\Psi_{1/2,+}^+)^{\dagger}{\mat v}_{\varphi}\Psi_{1/2,+}^+
-\frac e4 (\Psi_{1/2,-}^-)^{\dagger}{\mat v}_{\varphi}\Psi_{1/2,-}^-\nonumber\\
&-&\frac e4(\Psi_{-1/2,+}^-)^{\dagger}{\mat v}_{\varphi}\Psi_{-1/2,+}^-
-\frac e4(\Psi_{-1/2,-}^+)^{\dagger}{\mat v}_{\varphi}\Psi_{-1/2,-}^+\\
&=&-\frac{e\hbar}{4ma}\left[ \frac{1}{2}-\frac{1}{2 \cos\theta}\right ]
-\frac{e\hbar}{4ma}\left[ -\frac{1}{2}+\frac{1}{2 \cos\theta}\right ]\nonumber\\
&-&\frac{e\hbar}{4ma}\left[ -\frac{1}{2}-\frac{1}{2 \cos\theta}\right ]
-\frac{e\hbar}{4ma}\left[ \frac{1}{2}+\frac{1}{2 \cos\theta}\right ]=0
\end{eqnarray}
provided the SO strength is within the arrows in Fig. \ref{Fig1}, since outside the interval the computation
involves a different two-fold degenerate level. 
The result above applies to any normalized linear combination of
the fourfold ground states satisfying the time-reversal symmetry.
Charge current cancellation happens level by level and for all $\alpha$ 
values, rendering them zero as expected. The effect of an external magnetic 
field (playing the role of a time-reversal symmetry breaking field) 
and the ensuing charge persistent currents in the presence of a simple scalar potential and the spin-orbit interaction were examined in 
ref.\cite{Splettstoesser03}. The small scalar potential breaks the degeneracy
at the edge of the Brillouin zone that limits the charge persistent currents to a maximum value
for each band.

On the other hand, there can exist currents that don't break time reversal 
symmetry i.e. spin currents. In order to make an explicit calculation of 
the current, here we sidestep the consideration of a scalar potential 
and use the empty lattice approximation\cite{Ashcroft}. The spin current 
density, denoted as $\vec{\cal J}^a$  in order not to get confused 
with the charge current density $\vec J$, follows from the same approach as above. Note 
that the spin current $\vec{\cal J}^a$ is a tensor (two indices) while $\vec J$ is a vector.  We define now a
local ``spin velocity'' operator, properly symmetrized
\be
\vec {\cal J}^a=\frac{1}{2}\Psi^{\dagger}
\{ \vec {\mat v},{\mat s}^a\}\Psi,
\label{spin-velocity}
\ee 
where ${\mat s^a}=\hbar \bsigma^a/2$. The reader should note that the time reversal operation applied
to such an operator (in the absence of a magnetic field) reverses both the velocity and the spin, so the
current is not changed by such an operation, so this intuitive definition materializes our earlier expectation. 
\begin{figure} [th]
\vspace{0cm}
        \epsfxsize=12.0cm
        \begin{center}
        \mbox{\epsfbox{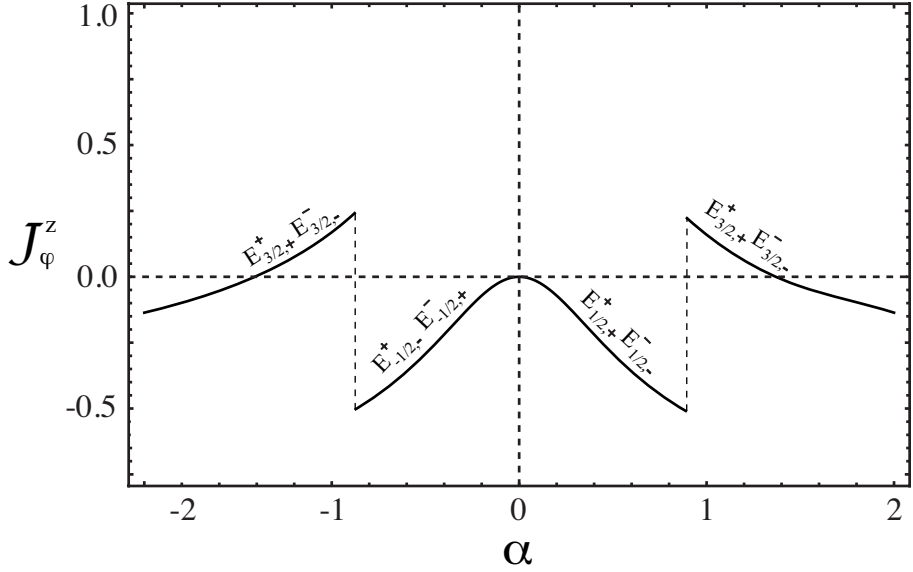}\qquad}
        \end{center}\vskip 0cm
        \caption{Spin current in units of $\hbar^2/4ma$ from the lowest lying degenerate level indicated in Fig.\ref{Fig1} as a function of $\alpha$. Between the arrows in Fig.\ref{Fig1} the lowest level is fourfold degenerate so there are four terms in the spin current. Outside that range, other two-fold degenerate level take over causing a jump in the spin current as indicated. The graph shows how the limit $\alpha=0$ indicates zero spin current because up down spin symmetry along with inversion symmetry are restored.}
        \label{Fig2}  
\end{figure}

The spin current along the ring will again involve the previously derived 
azimuthal velocity operator~(\ref{velocityoperator}). The general contribution for 
any single state involved in the current is given by
\begin{eqnarray}
{\cal J}_{\varphi}^a
&=&\frac{\hbar^2}{2ma}(\lambda n-1/2)[|A_{\lambda s}|^2\sigma_{11}^a+B_{\lambda s}^*A_{\lambda s} e^{-i\varphi}\sigma_{21}^a] \nonumber\\
&+&\frac{\hbar^2}{2ma}(\lambda n+1/2)[A_{\lambda s}^*B_{\lambda s} e^{i\varphi}\sigma_{12}^a+| B_{\lambda s}|^2\sigma_{22}^a]\nonumber\\
&-&\frac{\hbar}{2}\alpha \left (\cos\varphi~ \delta_{x,a}+\sin\varphi~ \delta_{y,a}\right ),
\end{eqnarray}
where $a=x,y,z$ and we have used notation from Eq. (\ref{eigenfunction}). 
Adding contributions from the four degenerate lowest lying 
levels (again the 
totally symmetric linear combination) 
with the spin orientation in the $z$ direction, one arrives at
\begin{eqnarray}
{\cal J}_{\varphi}^z&=&\frac 14(\Psi_{1/2,+}^+)^{\dagger}\frac{1}{2}\{{\mat v}_{\varphi},
{\mat s}^z\}\Psi_{1/2,+}^+
+\frac 14(\Psi_{1/2,-}^-)^{\dagger}\frac{1}{2}\{{\mat v}_{\varphi},{\mat s}^z\}\Psi_{1/2,-}^-\nonumber\\
&+&\frac 14(\Psi_{-1/2,+}^-)^{\dagger}\frac{1}{2}\{{\mat v}_{\varphi},
{\mat s}^z\}\Psi_{-1/2,+}^-
+\frac 14(\Psi_{-1/2,-}^+)^{\dagger}\frac{1}{2}\{{\mat v}_{\varphi},{\mat s}^z\}\Psi_{-1/2,-}^+\\
&=&\frac{\hbar^2}{8m a}\left (  \frac{1}{2}\cos\theta-\frac{1}{2} \right )
+\frac{\hbar^2}{8m a}\left (  \frac{1}{2}\cos\theta-\frac{1}{2} \right )
\nonumber\\
&-&\frac{\hbar^2}{8 m a}\left (  -\frac{1}{2}\cos\theta+\frac{1}{2} \right )
-\frac{\hbar^2}{8m a}\left ( - \frac{1}{2}\cos\theta+\frac{1}{2} \right ),\\
&=&\frac{\hbar^2}{4m a}\left (\cos\theta-1\right ).
\label{spincurrentzrashba}
\end{eqnarray}
Such an expression only accounts for $\alpha$ values in the range between the
arrows in Fig.\ref{Fig1}. Outside this range the energies $E_{3/2,+}^+, E_{3/2,-}^-$ are
the lowest energies, so we must compute the expectation of the spin current with such two-fold
degenerate eigenfunctions. The full spin current for the lowest lying levels is depicted in Fig.\ref{Fig2}.
One can see from the figure, that at $\alpha=0$ the spin currents vanish, as required by the
recovery of the inversion symmetry ($k\rightarrow -k$) and, independently, the spin inversion symmetry
that were intertwined in the presence of the SO interaction.
In the presence of a small scalar potential on the ring, the level crossing in Fig.\ref{Fig1}
gives way to level repulsion. The abrupt transition in dashed lines in Fig.\ref{Fig2} will then be
gradual and rounded\cite{PersistentCurrentOriginal}. Persistent spin currents have yet to measured
due to the lack of an appropriate probe.

\begin{figure} [th]
\vspace{0cm}
        \epsfxsize=11.0cm
        \begin{center}
        \mbox{\epsfbox{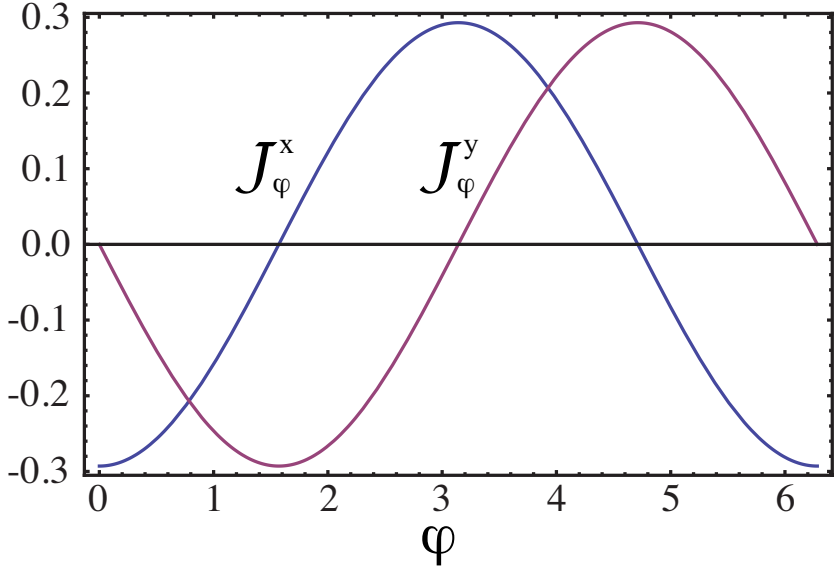}\qquad}
        \end{center}\vskip 0cm
        \caption{Spin currents for the $x,y$ polarizations in units of $\hbar^2/4ma$ from the lowest lying degenerate levels indicated in Fig.\ref{Fig1} as a function of $\varphi$ and for $\alpha=\frac{\hbar}{2ma}$. Between the arrows in Fig.\ref{Fig1}, the lowest level is fourfold degenerate so there are four terms in each of the spin currents.}
\label{Fig3}  
\end{figure}

The spin currents ${\cal J}_{\varphi}^{x,y}$ can also be readily computed along the lines above. In contrast to the spin current ${\cal J}_{\varphi}^{z}$, such currents for each of the degenerate states of a particular level, are not real valued. Nevertheless, the summation of all the degenerate  contribution 
renders a real spin currents which for the ground state is (in the spin-orbit strength range shown in Fig.\ref{Fig1})
\begin{eqnarray}
 {\cal J}_{\varphi}^x&=&\frac 12\hbar\alpha(\cos\theta-1)\cos\varphi,\nonumber\\
 {\cal J}_{\varphi}^y&=&\frac 12\hbar\alpha(\cos\theta-1)\sin\varphi.
 \end{eqnarray}
 Such expressions have a $\varphi$ dependence due to the precession of the spin around the $z$ axis. This means that a particular polarization rotates going through zero and then changing sign, making the whole spin current change sign while the wavevector is constant. Note that the spin persistent currents vanish correctly for $\alpha=0$ as expected. Figure~\ref{Fig3} shows the angular dependence for both spin current components for a value of $\alpha$ within the range shown in Fig.\ref{Fig1}.

%That spin currents arise is a natural consequence of the degeneracy of time 
%reversed states. Spin currents arise from adding currents with one spin 
%flowing clockwise with currents of the the opposite spin running 
%counterclockwise. This is exactly the two terms being added in the previous 
%equation. 
As with charge persistent currents one has to add all occupied 
levels. Observing figure \ref{Fig1} one sees that the first two levels add 
constructively while the third level subtracts from the previous two 
levels. One must then carefully add the corresponding currents taking
into account level crossings.
\subsection{The Dresselhaus SO interaction}
The Dresselhaus Hamiltonian of equation~(\ref{eqDresselhaus}) is 
another form of SO interaction. Using either of the
procedures shown in section 2, 
keeping only $p_\varphi$ terms, we express
$\bsigma_x p_x-\bsigma_y p_y$ as $-\bsigma_x\sin\varphi\ \!p_\varphi
-\bsigma_y\cos\varphi\ \!p_\varphi$ that we then symmetrize to get
\be\fl
\bar{rcl}
\V_{Dresselhaus}^{\hbox{\footnotesize ring}}
%\hbox{\rm Dresselhaus SO}
	&\!\!=\!\!
	&-\frac 12\beta(\bsigma_x\{\sin\varphi,p_\varphi\}
	+\bsigma_y\{\cos\varphi,p_\varphi\})\\ \\
	&\!\!=\!\!
	&i\hbar \beta a^{-1}[(\bsigma_x\sin\varphi+
	\bsigma_y\cos\varphi)\partial_\varphi+\frac 12(\bsigma_x\cos\varphi-
	\bsigma_y\sin\varphi)],
\ear
\ee
or we define 
the combination
\be \mat W=(\bsigma_x\sin\varphi+\bsigma_y\cos\varphi)i\partial_\varphi
+iA(\bsigma_x\cos\varphi-\bsigma_y\sin\varphi),\label{eqWbis}
\ee
and form the Hermitian conjugate
\be {\mat W}^\dagger
=(\bsigma_x\sin\varphi+\bsigma_y\cos\varphi)i\partial_\varphi
+i(1-A^*)(\bsigma_x\cos\varphi-\bsigma_y\sin\varphi),\label{eqWter}
\ee
where hermiticity requires $A^*=A=\frac 12$.

The method of Ref.~\cite{MMK02} applied to the Dresselhaus SO interaction
would require to express $\partial_x$ and $\partial_y$ in 
equation~(\ref{eqDresselhaus}) in terms of $\partial_\rho$ and 
$\partial_\varphi$, then to substitute $\rho$ by $a$ and $\partial_\rho$
by $-(2a)^{-1}$. It obviously leads to the same expression.

Now, completing the square as we have done in the Rashba case, we get
for the Dresselhaus SO interaction the more compact expression
\be
	\fl
	\H_{Dresselhaus}^{\hbox{\footnotesize ring}}
	=\frac{\hbar^2}{2ma^2}\left[\left(i\partial_\varphi
	\Un+\frac{ma\beta}{\hbar}(\bsigma_x\sin\varphi+\bsigma_y\cos\varphi)
	\right)^2-\left(\frac{ma\beta}{\hbar}\right)^2\Un\right].
\label{eqDresselhausRingSquare}
\ee
This expression is very similar to the Rashba Hamiltonian and only slight 
modifications are needed to calculate the spin current along the ring.
The energy levels are now given by
\be
E_{n,s}^{\lambda}=\frac{\hbar^2}{2ma^2}
\Bigl[
	\Bigl( 
		n-\frac{\lambda s}{2}\sqrt{1+4({ma\beta}/{\hbar})^2}
	\Bigr)^2-
	({ma\beta}/{\hbar}
	)^2
\Bigr],\label{energiesDresselhaus}
\ee
i.e. they are obtained from the Rashba eigenenergies via the substitution
$\alpha\to\beta$ and time reversal symmetry is preserved. 
The corresponding eigenspinors become
\begin{equation}
\Psi_{n,+}^+=e^{i n\varphi}\bpar{c}
\sin\frac{\theta}{2}\e^{i\varphi/2} \\ -i\cos\frac{\theta}{2}\e^{-i\varphi/2}
\epar,\\ \\
\Psi_{n,-}^+=e^{in\varphi}\bpar{c}
-\cos\frac{\theta}{2}\e^{i\varphi/2} \\ -i\sin\frac{\theta}{2}\e^{-i\varphi/2}
\epar
\end{equation}
\begin{equation}
\Psi_{n,+}^-=e^{-in\varphi}\bpar{c}
\sin\frac{\theta}{2}\e^{i\varphi/2} \\ -i\cos\frac{\theta}{2}\e^{-i\varphi/2}
\epar,\\ \\
\Psi_{n,-}^-=e^{-in\varphi}\bpar{c}
-\cos\frac{\theta}{2}\e^{i\varphi/2} \\ -i\sin\frac{\theta}{2}\e^{-i\varphi/2}
\epar
\end{equation}
where now $\cos\theta=1/\sqrt{1+4(m a \beta  /\hbar)^2}$.
When the SO coupling $\beta$ is not too strong, we recover the structure of a 
four-fold degenarate ground state and the spin current polarized in the $z$
direction takes the form
\be
{\cal J}_{\varphi}^z=\frac{\hbar^2}{4m a} (1-\cos\theta ).
\ee
Notice that the computed $z$ component of the spin current has the opposite sign to that of the Rashba
contribution (see (\ref{spincurrentzrashba})), so that given equal SO strengths for both type of interactions the two effects compete to produce a 
vanishing net spin current.

\subsection{Spin current and the non-Abelian gauge formalism}
The computed spin currents in the previous sections where based on intuitive extensions
of the very familiar charge current definition, i.e. the physical quantity being transported
times the velocity, along with the necessary symmetrization when the definition involves
two non-commuting operators. In this section we make the student realize that a
formal, well grounded approach can be used to define the currents based on regarding
them as conserved quantities following Noether's theorem. We will depart from a general Lagrangian
that will be identified with the corresponding ring Hamiltonian, derived above. New tensor
fields will arise, analogous to the gauge potential in electromagnetism but of a
non-abelian nature, that will by canonical relations, give us the full expressions of the currents 
in the theory.

The spin current calculated above is the equivalent of the paramagnetic 
current in the case of $U(1)$ gauge theory (see e.g. \cite{Berche09}).
One can ask about a possible equivalent to the diamagnetic contribution also
(such a contribution is called color diamagnetism 
by Tokatly~\cite{Tokatly,OurComment}).
The Lagrangian approach is very convenient to investigate this question.
To the Hamiltonian given in equation~(\ref{eqRashbaRingSquare}), we
associate a Lagrangian density ${\cal L}$ defined according to
\bea L&\equiv&\int ad\varphi\ \!{\cal L},\nonumber\\
&=&\langle\Psi|i\hbar\partial_t\Un-\H|\Psi\rangle,\nonumber\\
&=&\int ad\varphi\ \![i\hbar\Psi^\dagger\dot\Psi-\Psi^\dagger\H\Psi].
\eea
After an integration by parts, ${\cal L}$ follows,
\bea
\fl{\cal L}&\hskip-22mm =&\hskip-17mm i\hbar\Psi^\dagger\dot\Psi\nonumber\\
&&\hskip-17mm -\frac 1{2m}\left[
\Psi^\dagger(i\hbar a^{-1}\overleftarrow\partial_\varphi\Un-\frac 12gW^a_\varphi\bsigma_a))
\right]
\left[
(-i\hbar a^{-1}\overrightarrow\partial_\varphi\Un-\frac 12gW^a_\varphi\bsigma_a))\Psi
\right] \nonumber\\
&&\hskip-17mm +\frac 1{8m}g^2\Psi^\dagger(W^b_\varphi\bsigma_b)
(W^c_\varphi\bsigma_c)\Psi,\label{Lagrangian}
\eea
where the arrows above the partials indicate in which direction the derivative is taken and the non-Abelian gauge field
\be
\frac 12gW^a_\varphi\bsigma_a=m\alpha\bsigma_\rho,\label{nonAbelianW}
\ee
was introduced with $g=\hbar$ so that $W_{\varphi}^a$ has dimensions of $m\alpha/\hbar$. The $SU(2)$ gauge field is then proportional to the SO strength. The contraction on the internal 
index $a=x,y,z$ is understood.
From this Lagrangian, the spin current follows from derivatives with respect to the 
gauge field components,
\be
{\cal J}^a_\varphi=\frac{\partial {\cal L}}{\partial W^a_\varphi}
\ee
which has the correct dimension of a spin current.
A first contribution ${\cal J}_{(1)}$ to the spin current follows from
the second term at the r.h.s. of equation~(\ref{Lagrangian}),
\bea
{{\cal J}_{(1)}}^a_\varphi&=&-\frac 1{2m}\Psi^\dagger(-\frac 12\hbar\sigma_a)
(-i\hbar a^{-1}\partial_\varphi\Un-m\alpha\bsigma_\rho)\Psi
\nonumber\\
&&-\frac 1{2m}\Psi^\dagger
(i\hbar a^{-1}\partial_\varphi\Un-m\alpha\bsigma_\rho)
(-\frac 12\hbar\sigma_a)\Psi,
\eea
and can be explicitly calculated for all three spin components,
\bea
{{\cal J}_{(1)}}^x_\varphi&=&-\frac{i\hbar^2}{4ma}
[\Psi^\dagger\bsigma_x\partial_\varphi\Psi-
(\partial_\varphi\Psi^\dagger)\bsigma_x\Psi]
-\frac 12\hbar\alpha\Psi^\dagger\cos\varphi\Psi,\\
{{\cal J}_{(1)}}^y_\varphi&=&-\frac{i\hbar^2}{4ma}
[\Psi^\dagger\bsigma_y\partial_\varphi\Psi-
(\partial_\varphi\Psi^\dagger)\bsigma_y\Psi]
-\frac 12\hbar\alpha\Psi^\dagger\sin\varphi\Psi,\\
{{\cal J}_{(1)}}^z_\varphi&=&-\frac{i\hbar^2}{4ma}
[\Psi^\dagger\bsigma_z\partial_\varphi\Psi-
(\partial_\varphi\Psi^\dagger)\bsigma_z\Psi].
\eea
The ``diacolor'' contribution (linear in $\alpha$) mentioned above appears 
only for the $x$ and $y$ spin polarizations.
There is nevertheless another term in the Lagrangian, the third term
of the r.h.s. in equation~(\ref{Lagrangian}) which plays the role of
a {\it gauge symmetry breaking} term (or a mass term). It brings another
contribution ${\cal J}_{(2)}$ to the spin current,
\begin{eqnarray}
{{\cal J}_{(2)}}^a_\varphi&=&\frac 1{8m}g^2\Psi^\dagger \left (\sigma_a W^b_\varphi\sigma_b+ W^b_\varphi\sigma_b\sigma_a\right ) \Psi,\\
&=&\frac {1}{4}\hbar\alpha\Psi^\dagger \{\sigma_a, \sigma_{\rho}\} \Psi,
\end{eqnarray}
so
\begin{eqnarray}
{{\cal J}_{(2)}}^x_\varphi&=&\frac 12\hbar\alpha\Psi^\dagger \cos\varphi\Psi,\\
{{\cal J}_{(2)}}^y_\varphi&=&\frac 12\hbar\alpha\Psi^\dagger \sin\varphi\Psi,\\
{{\cal J}_{(2)}}^z_\varphi&=&0,
\end{eqnarray}
that exactly cancels the diacolor contribution in such a way that for all 
three spin polarizations only the ``paracolor'' contributions survive. 
Eventually the total spin current density, given by
\begin{equation}
{{\cal J}}^a_\varphi=-\frac{i\hbar^2}{4ma}
[\Psi^\dagger\bsigma_a\partial_\varphi\Psi-
(\partial_\varphi\Psi^\dagger)\bsigma_a\Psi],\\
\ee
is in agreement with the explicit calculations presented above for the 
$z-$polarization.
In the general case, due to the presence GSB term, it does not coincide 
with the usual definition of
the current in terms of the symmetrized spin-velocity 
product~(\ref{spin-velocity}). This property is not very well known and 
deserves special attention.

\section{Path integral on the ring and voltage quantization}
%\section{Voltage quantization}
Many questions in the context of mesoscopic rings concern phase effects.
When a perpendicular magnetic field is enclosed by the ring there occur
peculiar conditions to ensure a single-valued  wave function~\cite{ByersYang}. The well
known Aharonov-Bohm effect is a striking consequence of the peculiar phase
relations on a multiply connected structure such as a ring. In the presence
of SO interactions, phase relations are upgraded to spinor interferences
since the wavefunctions are now two component objects. Special attention must be paid when 
a SO interaction is present, since a new physical effect appears i.e. the spin precession. Spin
precession is a two component phase evolution in contrast to the one component phase
evolution in electromagnetism or $U(1)$ gauge theory. 

There is a simple qualitative picture, which consists in the interpretation
of the interaction term as similar to an effective Zeeman term,
$%\be
%\vec\bsigma\cdot(\vec {\rm E}\times\vec p)=
\sim -\vec\bsigma\cdot
\vec{\rm B}_{\rm eff.}
%\ee
$.
In the case of the Rashba SO interaction, the effective magnetic field
is 
\be
\vec{\rm B}^{\rm eff.}_{Rashba}=\alpha p_\varphi{\vec e}_\rho
\ee
and the axis around which the spin precesses rotates when the electron 
moves along
the ring.
%This would not be the case for a radial electric field as given in 
%equation (\ref{eqRadial}), since then
%\be
%\vec{\rm B}_{\rm eff.}^{Radial}=-\alpha p_\varphi{\vec e}_z.
%\ee
%The two situations differ strongly, since rotations around non collinear axis
%do not commute with each other, the first case poses problems of closure
%of the Pauli spinor except under some quantization prescriptions~\cite{Medina}%.

This condition appears when we consider the $SU(2)-$phase accumulation by the
spinor under transport along the loop~\cite{Medina}:
\be
\hbar^{-1}\oint\frac 12g\vec W^a\bsigma_a d\vec r=\frac{m\alpha}{\hbar}\int_0^{2\pi}
a\ \! d\varphi\ \!\bsigma_\rho.\label{Wilson}
\ee
This expression follows from the gauge formulation of the 
problem~\cite{ZhangYangMills,Medina}, and it is apparent for example in the
Rashba case in equation~(\ref{eqRashbaRingSquare}) where the spin of the
electron appears to be minimally coupled to the electric field through the
$SU(2)-$kinematic momentum
\be (-i\hbar a^{-1}\partial_\varphi)\Un-m\alpha\bsigma_\rho,
\ee
where $m\alpha\bsigma_\rho\equiv \frac 12g\vec W^a\bsigma_a$ is a
non-Abelian
$SU(2)-$gauge field introduced in equation~(\ref{nonAbelianW}).
When the operator $\exp\left ({i\hbar^{-1}
\oint\frac 12g\vec W^a\bsigma_a d\vec r}\right )$ 
constructed from equation~(\ref{Wilson}) is applied on 
$\Psi=\Pu|+\>_\rho+\Pd|-\>_\rho$, the
condition
\be
\cos{\textstyle\frac{2\pi ma\alpha}{\hbar}}\Un\Psi +i\bsigma_\rho
\sin {\textstyle\frac{2\pi ma\alpha}{\hbar}}\Psi=\Psi
\ee
follows to secure single-valued spinors. We thus obtain a quantization
condition
\be
\frac{2\pi ma\alpha}{\hbar}=2\pi\times{\tt integer}.
\ee
In the case of the standard SO interaction, $\alpha=|e|\hbar E/2m^2c^2$. 
This condition may be rewritten in terms of a typical voltage in the problem defined by $V=2\pi aE$ (take care 
here, the electrons on the ring move in a constant potential and the gate 
voltage applied externally to produce the electric field is a different 
quantity $V_{gate}$),
\be\frac{V}{4\pi mc^2/e}=
{\tt integer},\ee
where one can introduce a quantum of voltage, $V_0=4\pi mc^2/e$.

The same result also follows from more general arguments based on the
path integral formulation on the ring. As the student has learned in graduate quantum
mechanics, path integral allow for an alternative and illuminating derivation of quantum
mechanics generalizing  the action principle of classical mechanics. We will close
our description of ring physics in the presence of SO interactions by deriving in detail
its path integral and making contact with the quantization conditions derived above.

Let us consider again the Hamiltonian~(\ref{eqRashbaRingSquare})
%\be
%	\H_{Rashba}^{\hbox{\footnotesize ring}}
%	=\frac{\hbar^2}{2ma^2}\left[\left(i\partial_\varphi
%	\Un+\frac{ma\alpha}{\hbar}\bsigma_\rho
%	\right)^2-\left(\frac{ma\alpha}{\hbar}\right)^2\Un\right]
%\ee
and define the eigenstates $|n\rangle$ of the operator $i\partial_\varphi$,
$i\partial_\varphi|n\rangle=-n|n\rangle$, such that 
periodicity along the ring and normalization are 
satisfied, $\langle\varphi+2\pi|n\rangle=
\langle\varphi|n\rangle$ and 
$\int_0^{2\pi}d\varphi|\langle\varphi|n\rangle|^2=1$. We have 
\be
\langle\varphi|n\rangle=(2\pi)^{-1/2}e^{in\varphi},\quad n\in\mathbb{Z}.
\ee
In order to construct the amplitude for an electron to go from a 
particular state on the ring $|\varphi,\sigma;t\rangle$ to another similar state
at different time $|\varphi',\sigma';t'\rangle$, we construct the path 
integral in discretized 
time as
\be
{\cal A}(\varphi,\sigma;t\to\varphi',\sigma';t')=\prod_{i=1}^{N-1}
\langle\varphi_{i+1},\sigma_{i+1}|e^{-i\Delta t \H/\hbar}
|\varphi_i,\sigma_i\rangle,
\ee
with $\Delta t=t_{i+1}-t_i$. In the limit $\Delta t\to 0$, the evolution 
operator is expanded to linear order in $\Delta t$ and one has to evaluate 
matrix elements of the Hamiltonian between spin states. Expanding on the 
``plane wave'' basis, one has
\be i\partial_\varphi\Un|\varphi,\sigma\rangle = \sum_{n\in\mathbb{Z}}
(-n)(2\pi)^{-1/2}e^{-in\varphi}\Un|n,\sigma\rangle,
\ee
and the matrix element of the Hamiltonian reads as
\be
\fl\langle\varphi',\sigma'|\H|\varphi,\sigma\rangle=\frac{\hbar^2}{2ma^2}
\sum_{n\in\mathbb{Z}}\frac 1{2\pi}
[n^2\delta_{\sigma,\sigma'}-2n\frac{ma\alpha}{\hbar}
(\bsigma_\rho)_{\sigma,\sigma'}]e^{in(\varphi'-\varphi)}.
\ee
When we pass to the evolution operator, one has to evaluate
\be\langle\varphi',\sigma'|\Un-\frac{i\Delta t}\hbar\H|\varphi,\sigma\rangle
=\langle\varphi'|\varphi\rangle\delta_{\sigma,\sigma'}-
\frac{i\Delta t}\hbar\langle\varphi',\sigma'|\H|\varphi,\sigma\rangle,\ee
where we make use of the property
$\langle\varphi'|\varphi\rangle
=\sum_{n\in\mathbb Z}\frac 1{2\pi}e^{in(\varphi'-\varphi)}$ in order to 
factorize out the term $e^{in(\varphi'-\varphi)}$, and we exponentiate again
the Hamiltonian matrix element to get
\be\fl
\langle\varphi',\sigma'|\Un-\frac{i\Delta t}\hbar\H|\varphi,\sigma\rangle
=\sum_{n\in\mathbb Z}\frac 1{2\pi}
\left[e^{-\frac{i\hbar}{2ma^2}\Delta t[n^2\Un-2n\frac{ma\alpha}{\hbar}
\bsigmasmall_\rho]}\right]_{\sigma,\sigma'}
e^{in(\varphi'-\varphi)}.
\ee
Completing the square and introducing the classical variable 
$v_\varphi=a(\varphi'-\varphi)/\Delta t$, one eventually arrives at
\be
\langle\varphi_{i+1},\sigma_{i+1}|e^{-i\Delta t \H/\hbar}
|\varphi_i,\sigma_i\rangle={\rm const}\ \!\left[e^{
\frac i\hbar[{\frac 12} m(v_\varphi\Un-\alpha\bsigmasmall_\rho)^2]\Delta t}
\right]_{\sigma,\sigma'},
\ee
such that extended to the whole path, we get the symbolic expression
\be{\cal A}(\varphi,\sigma;t\to\varphi',\sigma';t')=\int\mathcal{D}\varphi(t)
{\bf T}\left[e^{\frac i\hbar\int{\frac 12} m(v_\varphi\Un-\alpha\bsigmasmall_\rho)^2dt}
\right]_{\sigma,\sigma'}
\label{amplitudeRashba},\ee
where {\bf T} is a super-operator ordering chronologically all operator 
products. We note that the path integral representation is only performed at the level of the space 
degrees of freedom and not for the spin variables for which we still  have an
evolution operator. Introducing spin coherent states, one could formulate 
path integrals where both coordinate and spin would be classical 
variables~\cite{PISpin}.
From the path integral, we can read a ``classical Lagrangian''
\be {\bf L}={\frac 12} m(v_\varphi\Un-\alpha\bsigma_\rho)^2.\ee
When we expand the square, we get three contributions to the spin precession.
If we consider the ``phase accumulation'' for a classical trajectory along 
a closed path over the ring (one round trip), 
the first term is simply the dynamical phase,
which is path dependent,
\be\frac 1\hbar\int_0^{2\pi}\frac 12 mv_\varphi a d\varphi
\ \! \delta_{\sigma,\sigma'}
=\frac{\pi mv_\varphi a}\hbar \delta_{\sigma,\sigma'}.\ee
It corresponds to the analogue of the optical phase.
The second term, the non-Abelian equivalent to the Aharonov-Bohm phase, is the
path independent (for closed paths) contribution discussed in the beginning 
of this section,
\be\frac 1\hbar \int_0^{2\pi} mv_\varphi
\alpha (\bsigma_\rho)_{\sigma,\sigma'}d t
=\frac{2\pi m\alpha (\bsigma_\rho)_{\sigma,\sigma'}a}{\hbar},\ee
or, using $\alpha$ and  $V$ defined above,
\be\frac V{2mc^2/e}(\bsigma_\rho)_{\sigma,\sigma'}.\ee
This is the counterpart of the famous $\Phi/\Phi_0$ phase in
$U(1)$ gauge theory.
The third term is an additional path-dependent contribution to the
phase which takes its origin in the initial GSB term of 
equation~(\ref{Lagrangian}),
\be\frac 1\hbar \int_0^{2\pi}\frac 12 m
\alpha^2(\bsigma_\rho^2)_{\sigma,\sigma'}
dt=\frac{\pi m\alpha^2 \delta_{\sigma,\sigma'}a}{\hbar v_\varphi}.
\label{quadraticterm}\ee
This last term  produces a phase variation which varies {\em quadratically} 
with the gate voltage applied.

Let us note the particular role of the GSB term in the Hamiltonian. In equation
(\ref{amplitudeRashba}), the transition amplitude is given by a path integral
where the action involves a term quadratic
in the gauge field. In the simpler case of a particle submitted to the 
$U(1)$-electromagnetic gauge field, the corresponding transition amplitude
would read as
\be
{\cal A}(\varphi;t\to\varphi';t')=\int\mathcal{D}\varphi(t)
e^{\frac i\hbar\int \left({\frac 12} mv_\varphi^2
+ev_\varphi A_\varphi\right)dt}, \ee
i.e. it would not include any quadratic term in the $U(1)$ gauge field.
The presence of such a quadratic term and its contribution
(\ref{quadraticterm}) to the phase accumulation is a direct consequence of the
GSB term in the SO case.

\section{Summary and Conclusions}

We have presented an overview of the analytical treatment and role of spin-orbit interactions in the
problem of mesoscopic ideal rings. For the benefit of the student, we have placed particular importance to the subtleties of
deriving the correct Hamiltonian to the problem when posing it in terms of cylindrical coordinates. Such details had been overlooked in the literature for some time until reference\cite{MMK02} clarified this point.  Besides the derivation given in the latter reference we have presented two other appealing approaches that
may offer a simpler procedure for more involved problems. Once the correct Hamiltonian was posed we set
out to explicitly derive the ground state properties in terms of the charge and spin currents as a function of the spin-orbit strength, pointing out the symmetries of the problem, and making such symmetries explicit both for the wavefunctions and the spectrum. Careful attention was paid to the correct limiting behavior as the spin-orbit interaction was sent to zero, the level degeneracies and all polarization components of the spin current for the lowest filling of the ring.

Using the Lagrangian formulation of the problems we formally derived the equilibrium currents, making a connection with recently reported {\it color currents}\cite{Tokatly}. Such an approach also permitted assessing  the role of the naturally ocurring gauge symmetry breaking term in the Rashba Hamiltonian, that is absent from the intuitive formulation of the current in terms of the velocity-spin anticommutator. The results point to a physical distinction of the gauge symmetry breaking contribution in the $x,y$ polarization components of the spin current. 

We finally addressed  topological considerations on the spin-orbit ring. A voltage quantization condition was derived due to uniqueness of the wavefunction around the ring. A path integral approach to the problem was also formulated in order to formally derive a quadratic in spin-orbit strength contribution to the geometrical phase around the ring.

Every one of the topics discussed has been an opportunity to convey to the student the subtleties of the quantum formulation of the  problem of spin transport on mesoscopic rings, detailing the calculations involved and making emphasis on physical insight, and not just bare technicalities.

There are many roads that begin from the material treated here. It is interesting to point out that the exact solutions derived are valid both for the limits: i) Slow rotation of the Rashba magnetic field with respect to a rapid precession of the spin and ii) Rapid rotation of the Rashba field in relation to the precession of the spin. The first is the adiabatic limit where Berry phase effects are dominant, while the latter relates to the sudden perturbation limit. Understanding physically the full range of such behaviors in the presence of a confining potential introduces the physics of lateral subbands that may render many interesting effects for mesoscopic spin-orbit rings as quantum circuit elements.

\ack We would like to acknowledge illuminating discussions with Victor Villalba. 
%%%%%%%%%%%%%%%%%%%%%%%%%%%%%%%%%%%%%%%%%%%%%%%%%%%%%%
%        REFERENCES
%%%%%%%%%%%%%%%%%%%%%%%%%%%%%%%%%%%%%%%%%%%%%%%%%%%%%%


\section*{References}
\vskip-12pt
\begin{thebibliography}{99}
\def\paper#1#2#3#4#5#6{{\rm #1}, {\it #6},  {#2}\ {\bf #3}, #4 (#5).}
\bibitem{Imry} Y. Imry, {\it Introduction to mesoscopic physics},
	Oxford University Press, Oxford 2002.
\bibitem{Konig06}
	\paper{M. K\"onig, A. Tschetschetkin, E.M. Hankiewicz, J. Sinova, 
	V. Hock, V. Daumer, M. Sch\"afer, C.R. Becker, H. Buhmann and 
	L.W. Molenkamp}{Phys. Rev. Lett.} {96} {076804} {2006}
	{Direct Observation of the Aharonov-Casher Phase}
%\bibitem{Nitta07}
%	\paper{J. Nitta and T. Bergsten}{New J. Phys.}{9}{341}{2007}
%	{Time reversal Aharonov-Casher effect using
%	Rashba spinâorbit interaction}
\bibitem{Kovalev07}
	\paper{A.A. Kovalev, M. F. Borunda, T. Jungwirth, L. W. Molenkamp 
	and J. Sinova}{Phys. Rev. B} {76} {125307} {2007}
	{Aharonov-Casher effect in a two-dimensional hole ring with 
	spin-orbit interaction}
\bibitem{Zivkovic08}
	\paper{M. Zivkovic, M. J\"a\"askel\"ainen, C.P. Search, and I. Djuric}
	{Phys. Rev. B} {77} {115306} {2008}
	{Sagnac rotational phase shifts in a mesoscopic electron 
	interferometer with spin-orbit interactions}
\bibitem{Frustaglia04}
	\paper{D. Frustaglia and K. Richter}
	{Phys. Rev. B} {69} {235310}{2004}
	{Spin interference effects in ring conductors subject to Rashba 
	coupling}
\bibitem{Harmer06}
	\paper{M. Harmer}{J. Phys. A} {39}{14329}{2006} 
	{Spin filtering on a ring with the Rashba Hamiltonian}
\bibitem{Hatano07}
	\paper{N. Hatano, R. Shirasaki and H. Nakamura}
	{Phys. Rev. A} {75} {032107}{2007}
	{Non-Abelian gauge field theory of the spin-orbit interaction and a 
	perfect spin filter}
\bibitem{Berche09}
	\paper{B. Berche, N. Bol\'\i var, A. L\'opez and E. Medina}
	{Cond. Matt. Phys.}{12}{707}{2009}
	{Gauge field theory approach to spin transport in a 2D electron gas}
\bibitem{Lopez09}
	\paper{A. L\'opez, E. Medina, N. Bol\'\i var and B. Berche}
	{J. Phys.: Condens. Matter}{22}{115303}{2010}
	{A perfect spin filtering device through Mach-Zehnder interferometry 
	in a GaAs/AlGaAs electron gas}
\bibitem{Recher07}
	\paper{P. Recher, B. Trauzettel A. Rycerz, Ya.M. Blanter, C.W.J. 
	Beenakker and A. F. Morpurgo}{Phys. Rev. B}{76} {235404} {2007}
	{Aharonov-Bohm effect and broken valley degeneracy in graphene rings}
\bibitem{Zhang06}
	\paper{Z.-Y. Zhang}
	{J. Phys. Condens. Matter}{18}{4101}{2006}
	{Spin accumulation on a one-dimensional mesoscopic
	Rashba ring}
\bibitem{Foldi06}
	\paper{P. F\"oldi, O. K\'alm\'an, M. G. Benedict and F.M. Peeters}
	{Phys. Rev. B}{73}{155325}{2006}
	{Quantum rings as electron spin beam splitters}
\bibitem{Citro06}
	\paper{R. Citro and F. Romeo}
	{Phys. Rev. B}{73}{233304}{2006}
	{Pumping in a mesoscopic ring with Aharonov-Casher effect}
\bibitem{Nitta09}
	\paper{J. Nitta, T. Bergsten, Y. Kunihashi and M. Kohda}
	{J. Appl. Phys.} {105}{122402}{2009}
	{Electrical manipulation of spins in the Rashba two dimensional 
	electron gas systems}
\bibitem{Molnar04}
	\paper{B. Moln\'ar, P. Vasilopoulos, and F. M. Peeters}
	{Appl. Phys. Lett.} {85}{612}{2004} 
	{Spin-dependent transmission through a chain of rings: Influence
	of a periodically modulated spinâorbit interaction strength 
	or ring  radius}
\bibitem{Zhang07}
	\paper{Z.-Y. Zhang}{J. Phys. Condens. Matter}{19}{236212}{2007} 
	{Andreev reflection in a mesoscopic hybrid
	four-terminal Rashba ring}
\bibitem{Foldi09}
	\paper{P. F\"oldi, M.G. Benedict, O. K\'alm\'an and F.M. Peeters}
	{Phys. Rev. B} {80}{165303}{2009}
	{Quantum rings with time-dependent spin-orbit coupling: Spintronic 
	Rabi oscillations and conductance properties}
\bibitem{Splettstoesser03}
	\paper{J. Splettstoesser, M. Governale and U. Z\"ulicke}
	{Phys. Rev. B} {68}{165341}{2003}
	{Persistent current in ballistic mesoscopic rings with Rashba 
	spin-orbit coupling}
\bibitem{Sun07}
	\paper{Q.-f. Sun, X.C. Xie and J. Wang}
	{Phys. Rev. Lett.} {98}{196801}{2007}
	{Persistent Spin Current in a Mesoscopic Hybrid Ring with 
	Spin-Orbit Coupling}
\bibitem{Huang09}
	\paper{G.-Y. Huang and S.-D. Liang}
	{EPL} {86}{67009}{2009}
	{Orbital magnetic phase and pure persistent spin current
	in spin-orbit coupling mesoscopic rings}
\bibitem{Kalman08}
	\paper{O. K\'alm\'an, P. F\"oldi, M.G. Benedict and F.M. Peeters}
	{Phys. Rev. B}{78}{125306}{2008}
	{Magnetoconductance of rectangular arrays of quantum rings}
\bibitem{Shelykh09}
	\paper{I.A. Shelykh, G. Pavlovic, D.D. Solnyshkov and G. Malpuech}
	{Phys. Rev. Lett.}{102}{046407}{2009}	
	{Proposal for a Mesoscopic Optical Berry-Phase Interferometer}
\bibitem{Foldi05}
	\paper{P. F\"oldi B. Moln\'ar, M.G. Benedict and F. M. Peeters}
	{Phys. Rev. B} {71}{033309}{2005}
	{Spintronic single-qubit gate based on a quantum ring with 
	spin-orbit interaction}
\bibitem{Sakurai} J. J. Sakurai, {Modern Quantum Mechanics},  Addison Wesley, Reading 1994.
\bibitem{BjorkenDrell} J.D. Bjorken and S.D. Drell, {\it Relativistic quantum 
	mechanics}, Mac Graw Hill, New-York 1964.
\bibitem{Merzbacher} E. Merzbacher, {\it Quantum Mechanics}, John Wiley and Sons, Inc., New York 1970.
\bibitem{Rashba}\paper{E.I. Rashba}{Sov. Phys. Solid State}{2}{1109}{1960}
	{Properties of semiconductors with an extremum loop}
\bibitem{BychkovRashba} \paper{Y.A. Bychkov and E.I. Rashba}
	{JETP Lett.}{39}{78}{1984}
	{Properties on 2D electron-gas with  lifted spectrum degeneracy}
\bibitem{Ganichev}
	\paper{S. D. Ganichev, V.V. Bel'kov, L.E. Golub, E.L. Ivchenko, 
	Petra Schneider, S. Giglberger, J. Eroms, J. De Boeck, G. Borghs, 
	W. Wegscheider, D. Weiss, and W. Prettl}
	{Phys. Rev. Lett.}{92}{256601}{2004}
	{Experimental separation of Rashba and Dresselhaus spin 
	splittings in semiconductor quantum wells}	
\bibitem{EngelRashbaHalperin} H.A. Engel, E. I. Rashba and B.I. Halperin,
   	{\it Theory of Spin Hall Effects in Semiconductors}, 
	in Handbook of Magnetism and Advanced Magnetic Materials, 
	H. Kronm\"uller and S. Parkin, eds. 
	(John Wiley \& Sons, Chichester, UK, 2007) 
	Vol. 5, pp 2858-2877. Cond-mat/0603306
\bibitem{winkler} R. Winkler, 
	{\it Spin-Orbit Coupling Effects in Two Dimensional Electron 
	and Hole Systems}, (Springer) 2003.
\bibitem{Rashba06} E.I. Rashba, {\it Semiconductors spintronics: Progress
	and Challenges}, arXiv:cond-mat/0611194
%\bibitem{Lipparini} E. Lipparini, {\it Modern many-particle physics},
%	World Scientific, Singapore 2008.
\bibitem{MMK02} \paper{F.E. Meijer, A.F. Morpurgo and T.M. Klapwijk}
	{Phys. Rev. B}{66}{033107}{2002}{One-dimensional ring in the presence
	of Rashba spin-orbit interaction: Derivation of the correct 
	Hamiltonian}
\bibitem{Paz} \paper{G. Paz}{Eur. J. Phys.}{22}{337}{2001}{On the connection
	between the radial momentum operator and the Hamiltonian in $n$ 
	dimensions}
\bibitem{MolnarPeetersVasilopoulos} 
	\paper{B. Moln\'ar, F. M. Peeters and P. Vasilopoulos}
	{Phys. Rev. B}{69}{155335}{2004}
	{Spin-dependent magnetotransport through a ring due to 
	spin-orbit interaction}
\bibitem{PeskinSchroder} M. Peskin and D. Shroeder, 
	{\it An Introduction to Quantum Field Theory}, 
	(Boulder, Westview Press) 1996.
\bibitem{PersistentCurrentOriginal} \paper{M. Buttiker, Y. Imry, and R. 
	Landauer}{Phys. Lett.}{96A}{365}{1983}
	{Josephson behavior in small normal one dimensional rings}
\bibitem{Mohanty} 
	\paper{P. Mohanty}
	{Ann. Phys. (Leipzig)}{8}{549}{1999}
	{Persistent currents in normal metals}	
\bibitem{Ashcroft}N. Ashcroft, D. Mermin, 
	{\it Solid State Physics}, (Holt, Rinehart and Winston) 1976.
\bibitem{Tokatly} \paper{I.V. Tokatly}{Phys. Rev. Lett.}{101}{106601}{2008}
	{Equilibrium Spin Currents: Non-Abelian Gauge Invariance and Color 
	Diamagnetism in Condensed Matter}
\bibitem{OurComment}A. L\'opez, E. Medina, N. Bol\'\i var and B. Berche,
	{\it Comment on Equilibrium Spin Currents: Non-Abelian Gauge 
	Invariance and Color Diamagnetism in Condensed Matter}
	arxiv/0902.46352009 (2009).
\bibitem{ByersYang}\paper{N. Byers and C.N. Yang}{Phys. Rev. Lett.}{7}{46}
	{1961}{Theoretical considerations concerning quantized magnetic 
	flux in superconducting cylinders}
\bibitem{Medina} \paper{E. Medina, A. L\'opez, and B. Berche}
	{Europhys. Lett.}{83}{47005}{2008}{Gauge symmetry breaking and 
	topological quantization
	for the Pauli Hamiltonian}
\bibitem{ZhangYangMills}  
	\paper{P.Q. Jin,  Y.Q. Li and  F.C. Zhang} 
	{J. Phys. Math Gen.} {39}{7115}{2006} 
	{$SU(2) \times U(1)$ unified theory for charge, orbit and spin currents}
\bibitem{PISpin} \paper{P. Lucignano, D. Giuliano, A. Tagliacozzo}
	{Phys. Rev. B}{76}{045324}{2007}{Quantum Rings with Rashba spin orbit coupling: a path integral approach}


\end{thebibliography}
\end{document}